# Deep Neural Networks for Accurate Predictions of Garnet Stability

Weike Ye[1], Chi Chen[2], Zhenbin Wang[2], Iek-Heng Chu[2] & Shyue Ping Ong[2]

**Predicting the stability of crystals is one of the central problems in materials science. Today, density functional theory (DFT)[1] calculations are the computational tool of choice to obtain energies of crystals with quantitative accuracy. Despite algorithmic[2] and computing advances, DFT calculations remain comparatively expensive and scale poorly with system size. Here we show that deep neural networks[3] utilizing just two descriptors – the Pauling electronegativity[4] and ionic radii[5] – can predict the DFT formation energies of $C_3A_2D_3O_{12}$ garnets with extremely low mean absolute errors of 7–8 meV/atom, an order of magnitude improvement over previous machine learning models and well within the limits of DFT accuracy. Further extension to mixed garnets with little loss in accuracy can be achieved using a binary encoding scheme that introduces minimal increase in descriptor dimensionality. Our results demonstrate that generalizable deep-learning models for quantitative crystal stability prediction can be built on a small set of chemically-intuitive descriptors. Such models provide the means to rapidly transverse vast chemical spaces to accurately identify stable compositions, accelerating the discovery of novel materials with potentially superior properties.**

[1] Department of Chemistry and Biochemistry, University of California San Diego, 9500 Gilman Dr, Mail Code 0303, La Jolla, CA 92093-0448, United States

[2] Department of NanoEngineering, University of California San Diego, 9500 Gilman Dr, Mail Code, 0448, La Jolla, CA 92093-0448, United States

The formation energy of a crystal is a key metric of its stability and synthesizability. It is typically defined relative to constituent unary/binary phases ($E_f$) or the stable linear combination of competing phases in the phase diagram ($E_{hull}$, or energy above convex hull)[6]. In recent years, machine learning (ML) models trained on DFT calculations have garnered widespread interest as a means to scale quantitative predictions of materials properties[7–11], including energies of crystals. However, previous efforts at predicting $E_f$ or $E_{hull}$ of crystals[9,12,13] using ML models have yielded mean absolute errors (MAEs) of > 100 meV/atom, falling far short of the necessary accuracy for useful crystal stability predictions. Approximately 90% of the crystals in the Inorganic Crystal Structure Database (ICSD) have $E_{hull}$ < 70 meV/atom[14], and the errors of DFT-calculated formation energies of ternary oxides from binary oxides relative to experiments are ~ 24 meV/atom.[15]

We have approached the crystal stability prediction problem by using artificial neural networks (ANNs), i.e., algorithms that are loosely modelled on the mammalian brain, to quantify well-established chemical intuition. The Pauling electronegativity and ionic radii guide much of our understanding about the bonding and stability of crystals today, for example, in the form of Pauling's five rules[16] and the Goldschmidt tolerance factor for perovskites[17]. Though these rules are highly qualitative, their success points to the potential existence of a direct relationship between crystal stability and these descriptors.

To probe these relationships, we have chosen, as our model system, the garnets, a large family of crystals with widespread technological applications such as luminescent materials for solid-state lighting[18] and lithium superionic conductors for rechargeable lithium-ion batteries[19,20]. Garnets have the general formula $C_3A_2D_3O_{12}$, where C, A and D denote the three cation sites with Wyckoff symbols 24*c* (dodecahedron), 16*a* (octahedron) and 24*d* (tetrahedron), respectively, in the prototypical cubic $Ia\bar{3}d$ garnet crystal shown in Fig. 1. The distinct coordination environments of the three sites result in different minimum ionic radii ratios (and hence, species preference) according to Pauling's first rule.

We start with the hypothesis that the formation energy $E_f$ of a $C_3A_2D_3O_{12}$ garnet is some unknown function $f$ of the Pauling electronegativities ($\chi$) and Shannon ionic radii ($r$) of the species in the C, A and D sites, i.e.,

$$E_f = f(\chi_C, r_C, \chi_A, r_A, \chi_D, r_D) \qquad (1)$$

Here, we define $E_f$ as the change in energy in forming the garnet from binary oxides with elements in the same oxidation states. Using the $Ca_3Al_2Si_3O_{12}$ garnet (grossular) as an example, $E_f$ is given by the energy of the reaction: $3CaO + Al_2O_3 + 3SiO_2 \rightarrow Ca_3Al_2Si_3O_{12}$. This choice of definition of $E_f$ is motivated by three reasons. First, binary oxides are frequently used as synthesis precursors. Second, our definition ensures that garnets that share elements in the same oxidation states have $E_f$ that are referenced to the same binary oxides. Finally, this $E_f$ definition minimizes DFT errors associated with the self-interaction error in semi-local functionals, which is incompletely cancelled out in redox reactions[21]. In contrast, $E_{hull}$ is a poor target metric for a ML model as it is defined with respect to the linear combination of stable phases at the $C_3A_2D_3O_{12}$ composition in the C-A-D-O phase diagram, which can vary unpredictably even for highly similar chemistries.

Based on the universal approximation theorem[22], we may model the unknown function $f(\chi_C, r_C, \chi_A, r_A, \chi_D, r_D)$, which is clearly non-linear using a feed-forward artificial neural network (ANN), as depicted in Fig. 2. The loss function and metric are chosen to be the mean squared error (MSE) and MAE, respectively. We will denote the architecture of the ANN using $n^i$-$n^{[1]}$-$n^{[2]} - \cdots - 1$, where $n^i$ and $n^{[l]}$ are the number of neurons in the input and $l^{\text{th}}$ hidden layer, respectively.

We developed an initial ANN model for unmixed garnets, i.e., garnets with only one type of species each in C, A and D. The training and test data comprising 704 unmixed garnets were generated by performing full DFT relaxation and energy calculations (see Methods) on all charge-neural combinations of allowed species on the C, A and D sites[23]. Using a training:test ratio of 80:20 and 50-fold repeated random sub-sampling cross validation, we find that a 6-25-1 ANN architecture yields a small root mean square error (RMSE) of 11 meV/atom, as well as the smallest standard deviation in the RMSE among the 50 sub-samples. We selected the model with the lowest $L^2$ norm of weights to minimize the risk of over-fitting. The training and test MAEs for the optimized 6-25-1 model are ~ 7–8 meV/atom (Fig. 3a), more than an order of magnitude lower than the ~ 100 meV/atom achieved in previous ML models.[9,12,13] For comparison, the error in the DFT $E_f$ relative to experimental values is around 14 meV/atom. Similar RMSEs are obtained for deep neural network (DNN) architectures containing two hidden layers , indicating that a single-hidden-layer architecture is sufficient to model the relationship $E_f$ and the descriptors.

To extend our model to mixed garnets, i.e., garnets with more than one type of species in the C, A and D sites, we explored two alternative approaches – one based on averaging of descriptors, and another based on expanding the number of descriptors to account for the effect or species ordering. The training and test data for mixed garnets were created using the same species pool, but allowing two species to occupy one of the sites. Mixing on the A sites was set at a 1:1 ratio, and that on the C and D sites was set at a 2:1 ratio, generating garnets of the form $C_3A'A''D_3O_{12}$ (215 compositions), $C'C_2''A_2D_3O_{12}$ (494 compositions) and $C_3A_2D'D_2''O_{12}$ (116 compositions). For each composition, we calculated the energies of all symmetrically distinct orderings within a single primitive unit cell of the garnet. All orderings must belong to a subgroup of the $Ia\bar{3}d$ garnet space group.

In the first approach, we characterized each C, A or D site using weighted averages of the ionic radii and electronegativities of the species present in each site, given by the following expressions (see Methods):

$$r_{avg} = xr_X + (1-x)r_Y \qquad (2)$$

$$\chi_{avg} = \chi_O - \sqrt{x\,(\chi_X - \chi_O)^2 + (1-x)(\chi_Y - \chi_O)^2} \qquad (3)$$

where X and Y are the species present in a site with fraction $x$ and $(1-x)$, respectively, and O refers to the element oxygen. The implicit assumption in this "averaged" ANN model is that species X and Y are completely disordered, i.e., different orderings of X and Y result in negligible DFT energy differences.

Using the same 6-25-1 ANN architecture, we fitted an "averaged" model using the energy of the *ground state ordering* of the 704 unmixed and 825 mixed garnets. We find that the training and test MAEs of the optimized model are 14 and 17 meV/atom, respectively (Fig 3b). These MAEs are about double that of the unmixed ANN model, but still significantly lower (~ 5 times) than prior ML models[9,12,13], and comparable to the error of the DFT $E_f$ relative to experiments. The larger MAEs may be attributed to the fact that the effect of species orderings on the crystal energy is not accounted for in this "averaged" model.

In the second approach, we undertook a more ambitious effort to account for the effect of species orderings on crystal energy. Here, we discuss the results for species mixing on the C site only, for which the largest number of computed compositions and orderings is available. For 2:1 mixing, there are 20 symmetrically distinct orderings within the primitive garnet cell, which can be encoded using a 5-bit binary array $[b_0, b_1, b_2, b_3, b_4]$. This binary encoding scheme is significantly more compact that the commonly-used one-hot encoding scheme, and hence, minimizes the increase in the descriptor dimensionality. We may then modify Eqn. 1 as follows:

$$E_f = f(\chi_{C'}, r_{C'}, \chi_{C''}, r_{C''}, \chi_A, r_A, \chi_D, r_D, b_0, b_1, b_2, b_3, b_4) \qquad (4)$$

where the electronegativities and ionic radii of both species on the C sites are explicitly represented. In contrast to the "averaged" model, we now treat the 20 ordering-$E_f$ pairs at each composition as distinct data points. Each unmixed composition was also included as 20 data points with different binary encodings but the same $E_f$. We find that a two-hidden-layer deep neural network (DNN) is necessary to model this more complex composition-ordering-energy relationship. The final optimized 13-22-8-1 model exhibits training and test MAEs of ~ 8 meV/atom (Fig 3b). The comparable MAEs between this extended DNN model and the unmixed ANN model is clear evidence that the DNN model has successfully captured the additional effect of orderings on $E_f$. Furthermore, the standard deviation of $E_f$ across orderings for *unmixed* garnets predicted using the extended DNN model is only ~ 2 meV/atom, i.e., different orderings have negligible effect on the $E_f$ when only one species is present. Finally, similar MAEs can be achieved for A and D site mixing (Fig 3b) using the same approach.

While $E_f$ is a good target metric for a predictive ANN model, the stability of a crystal is ultimately characterized by its $E_{hull}$. Using the predicted $E_f$ from our DNN models and pre-calculated DFT data from the Materials Project[24], we have computed $E_{hull}$ by constructing the 0 K C-A-D-O phase diagrams. From Fig. 4, we may observe that the extended DNN models can achieve a > 77% accuracy in classifying stable/unstable unmixed garnets at a strict $E_{hull}$ threshold of 0 meV/atom, and rises rapidly with increasing threshold. Similarly, high classification accuracies of greater than 90% are achieved for all three types of mixed garnets.

Given the great flexibility of the garnet prototype in accommodating different species, there are potentially millions of undiscovered compositions. Even using our restrictive protocol of single-site mixing in specified ratios, more than 10,000 mixed garnet compositions can be generated, of which 7287 are predicted to have $E_{hull}$ < 30 meV/atom, i.e., potentially synthesizable. A web application that computes $E_f$ and $E_{hull}$ for any garnet composition using the optimized DNNs has been made publicly available for researchers. More importantly, we have shown that DNNs can quantify the chemically-intuitive relationship between descriptors such as the Pauling electronegativity and ionic radii, and crystal energy. The successful extension to mixed garnets belonging to subgroups of $Ia\bar{3}d$ indicates that this is an approach that is likely to be generalizable to other crystal structure prototypes.

## Methods Summary

**Density functional theory (DFT) calculations.** All DFT calculations were performed using Vienna *ab initio* simulation package (VASP) within the projector augmented wave approach[25,26]. Calculation parameters were chosen to be consistent with those used in the Materials Project, an open database of pre-computed energies for all known inorganic materials[24]. The Perdew-Burke-Ernzehof generalized gradient approximation exchange-correlation functional[27] and a plane-wave energy cut-off of 520 eV were used. Energies were converged to within $5 \times 10^{-5}$ eV, and all structures were fully relaxed. For mixed garnets, symmetrically distinct orderings within the 80-atom primitive unit cell were generated using the Python Materials Genomics package.[28]

**Training of artificial neural networks (ANNs).** Training of the ANNs was carried out using the Adam optimizer[29] at a learning rate of 0.2, with the mean square error of $E_f$ as the loss metric. For each architecture, we ran 50 iterations with a random 80:20 split of training:test data, i.e., repeated random sub-sampling cross validation.

**Electronegativity averaging.** Pauling's definition of electronegativity is based on an "additional stabilization" of a heteronuclear bond X-O compared to average of X-X and O-O bonds, as follows.

$$(\chi_X - \chi_O)^2 = E_d(XO) - \frac{E_d(XX) + E_d(OO)}{2}$$

where $\chi_X$ and $\chi_O$ are the electronegativities of species X and O, respectively, and $E_d$ is the dissociation energy of the bond in parentheses. Here, O refers to oxygen.

For a disordered site containing species X and Y in the fractions $x$ and $(1-x)$, respectively, we obtain the following:

$$\left(\chi_{X_xY_{1-x}}-\chi_O\right)^2 = xE_d(\text{XO}) + (1-x)E_d(\text{YO}) - \frac{xE_d(\text{XX}) + (1-x)E_d(\text{YY}) + E_d(\text{OO})}{2}$$

$$= x(\chi_X-\chi_O)^2 + (1-x)(\chi_Y-\chi_O)^2$$

We then obtain the effective electronegativity for the disordered site as follows:

$$\chi_{X_xY_{1-x}} = \chi_O - \sqrt{x(\chi_X-\chi_O)^2 + (1-x)(\chi_Y-\chi_O)^2}$$

## Acknowledgements

This work is supported by the Samsung Advanced Institute of Technology (SAIT)'s Global Research Outreach (GRO) Program. The authors also acknowledge data and software resources provided by the Materials Project, funded by the U.S. Department of Energy, Office of Science, Office of Basic Energy Sciences, Materials Sciences and Engineering Division under Contract No. DE-AC02-05-CH11231 : Materials Project program KC23MP, and computational resources provided by Triton Shared Computing Cluster (TSCC) at the University of California, San Diego, the National Energy Research Scientific Computing Centre (NERSC), and the Extreme Science and Engineering Discovery Environment (XSEDE) supported by National Science Foundation under Grant No. ACI-1053575. The authors would also like to express their gratitude to Professors Darren Lipomi and David Fenning from the University of California, San Diego, and Dr Anubhav Jain from Lawrence Berkeley National Laboratory for helpful comments on the manuscript.

## Author Contributions

S.P.O, W.Y and C.C proposed the concept. W.Y carried out the calculations and analysis with the help from C.C., Z.W. and I.C. W.Y. prepared the initial draft of the manuscript. All authors contributed to the discussions and revisions of the manuscript.

## Author Information

The authors declare no competing financial interests. Correspondence and requests for materials should be addressed to ongsp@eng.ucsd.edu.

**Figure legends**

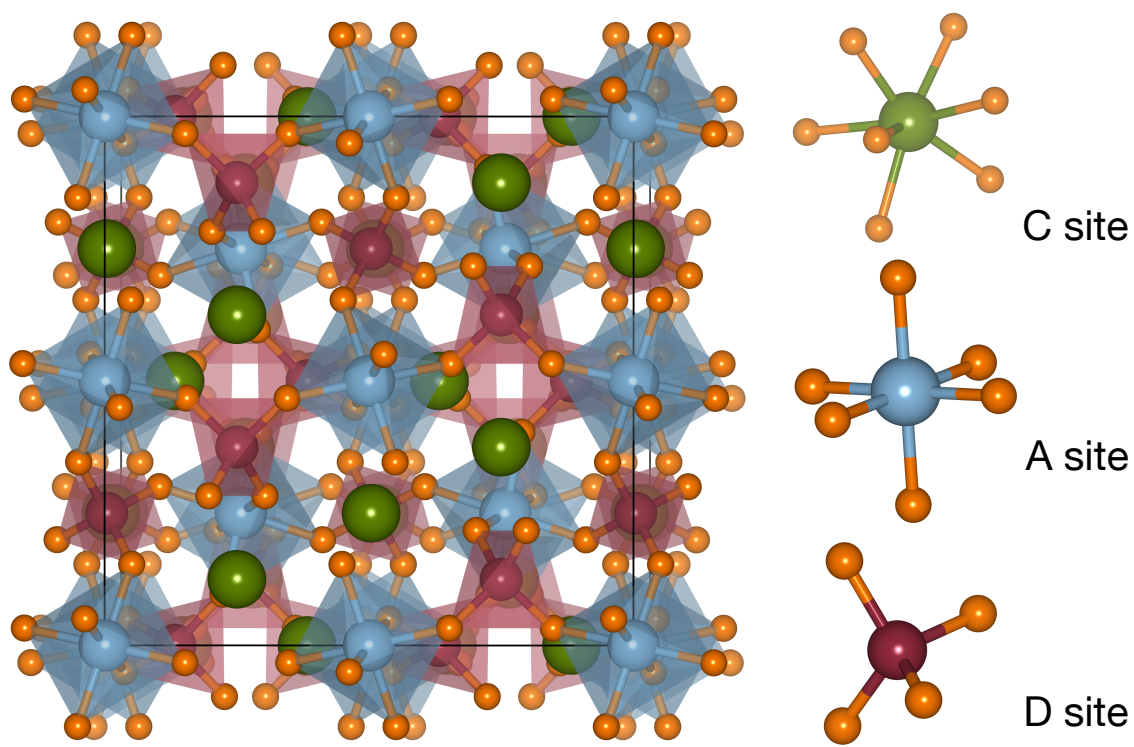

**Figure 1 | Crystal structure of $Ia\overline{3}d$ $C_3A_2D_3O_{12}$ garnet prototype.** Green (C), blue (A) and red (D) spheres are atoms in the 24*c* (dodecahedron), 16*a* (octahedron) and 24*d* (tetrahedron) sites, respectively. Orange spheres are oxygen atoms.

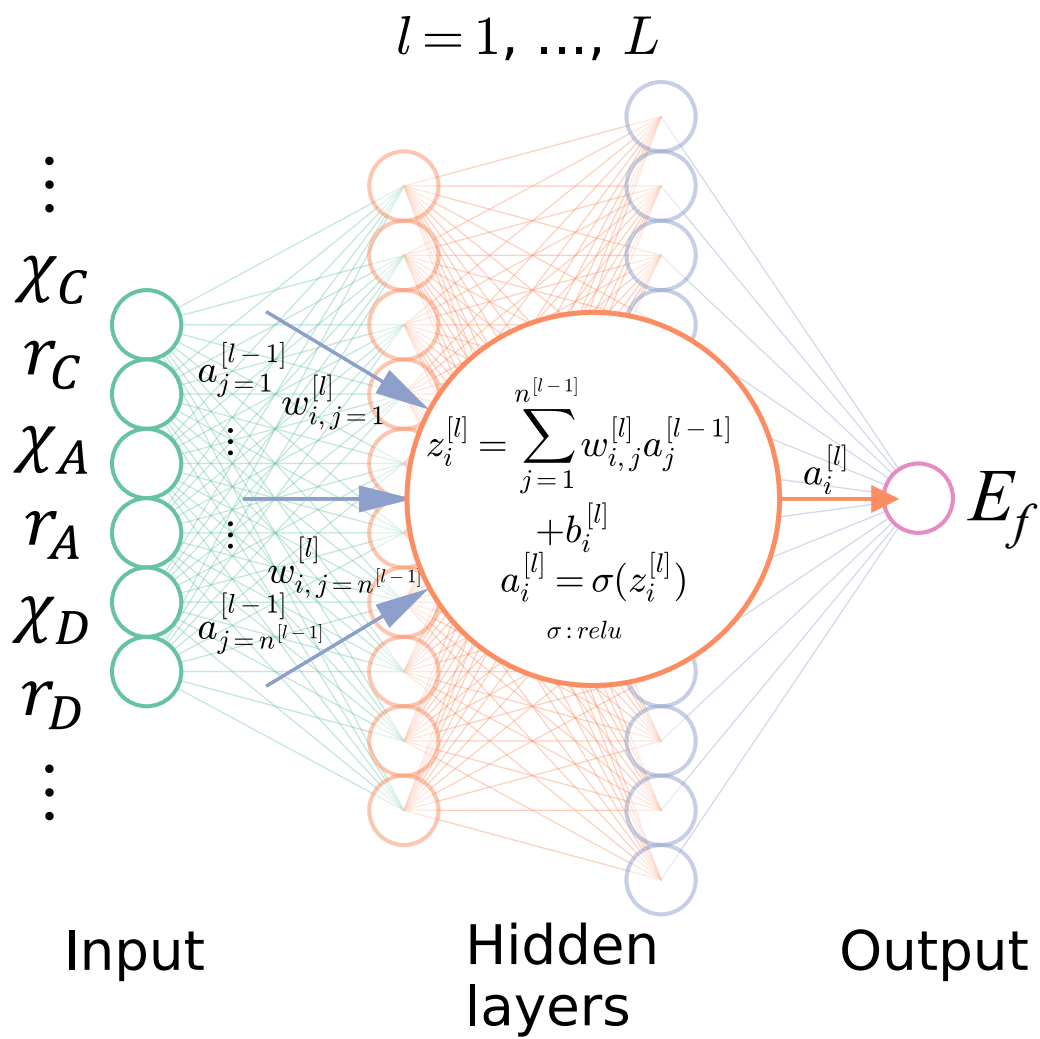

**Figure 2 | General schematic of the artificial neural network.** The artificial neural network (ANN) comprises an input layer of descriptors (the Pauling electronegativity and ionic radii on each site), followed by a number of hidden layers, and finally an output layer ($E_f$). The large circle in the centre shows how the output of the $n_{th}$ neuron in $l_{th}$ layer, $a_i^{[l]}$, is related to the received inputs from $(l-1)_{th}$ layer $a_j^{[l-1]}$. $w_{i,j}^{[l]}$ and $b_i^{[l]}$ denote the weight and bias between the $j_{th}$ neuron in $(l-1)_{th}$ layer and $i_{th}$ neuron in $l_{th}$ layer. σ is the activation function (rectified

linear unit in this work). The ANN models were implemented using Keras[30] deep learning library with the Tensorflow[31] backend.

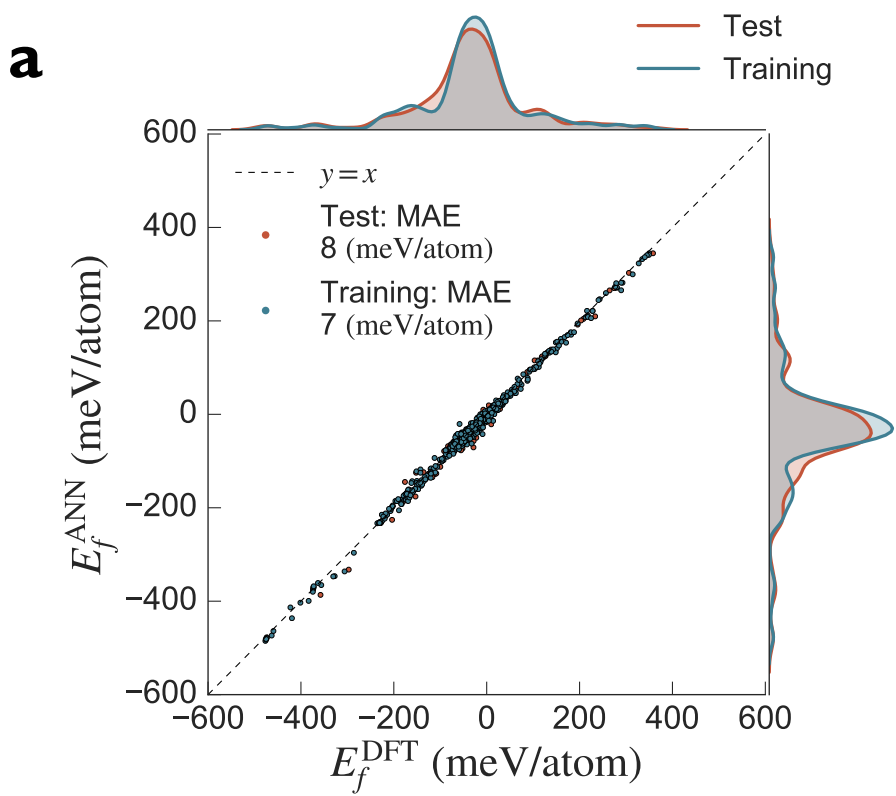

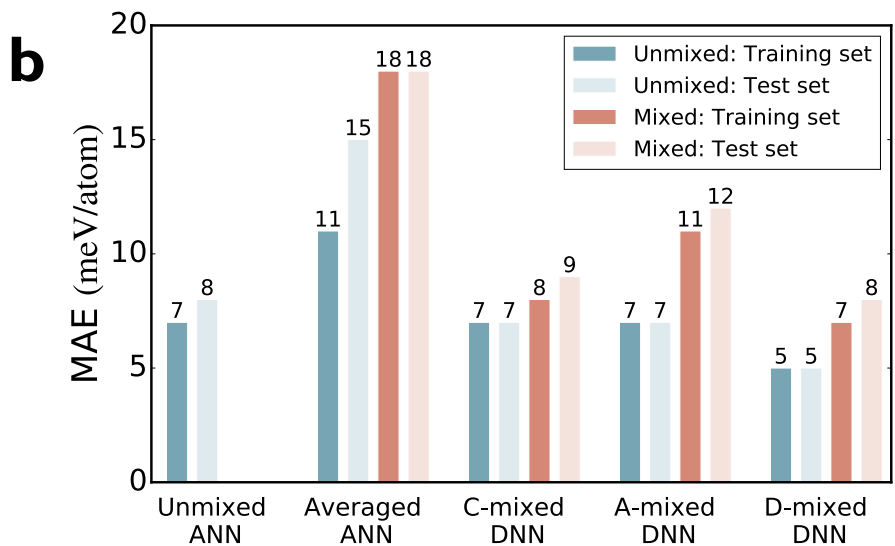

**Figure 3 | Performance of artificial neural network (ANN) models. a.** Plot of $E_f^{ANN}$ against $E_f^{DFT}$ of unmixed garnets for optimized 6-25-1 ANN model. The histograms at the top and right show that both the training and test sets contain a good spread of data across the entire energy range of interest. Low mean absolute errors (MAEs) in $E_f$ of 7 and 8 meV/atom are observed for the training and test set, respectively. **b.** MAEs in $E_f$ for training and test sets of all models. The C-, A- and D-mixed deep neural networks (DNNs) have similar training and test MAEs as the unmixed ANN model, indicating that the neural network has learned the effect of orderings on $E_f$. Each C, A and D composition has 20, 7, and 18 distinct orderings, respectively, which are encoded using 5-bit, 3-bit and 5-bit binary arrays, respectively.

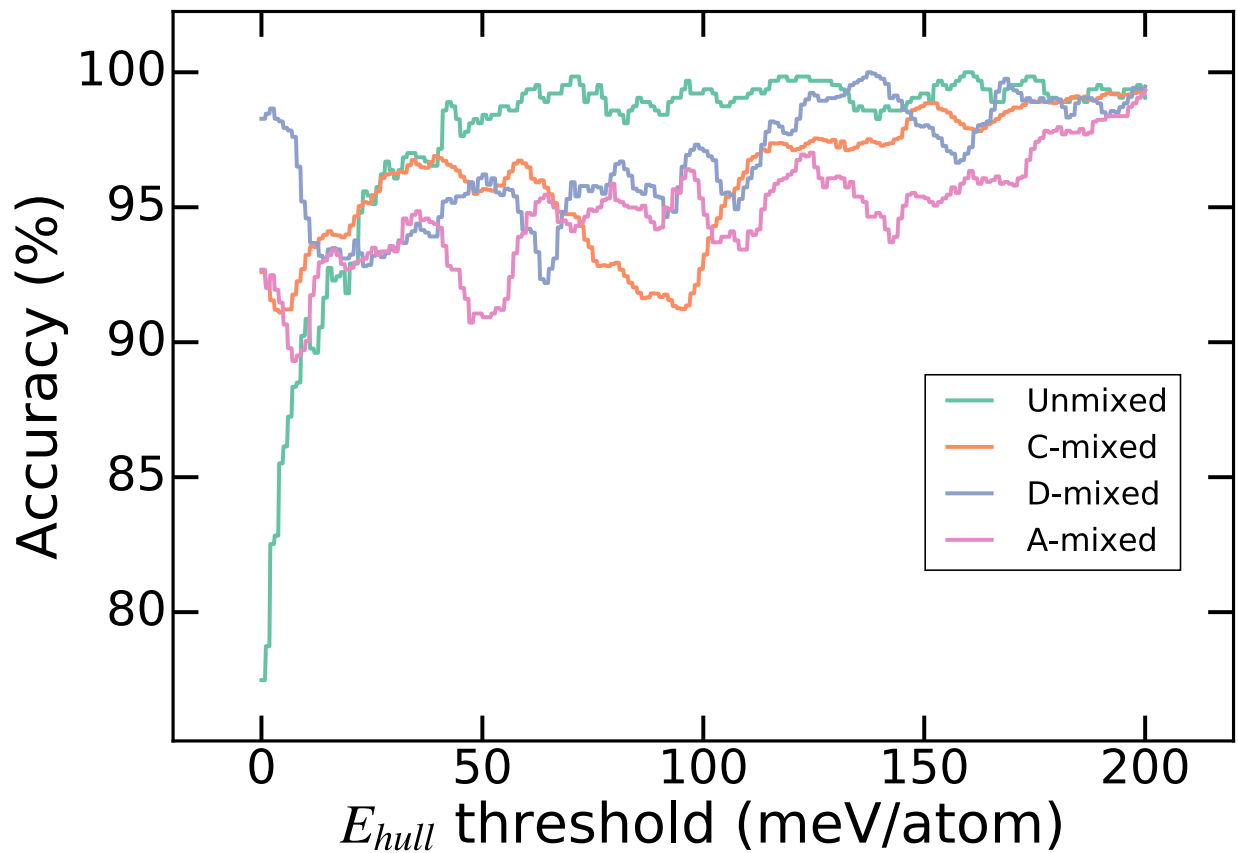

**Figure 4 | Accuracy of stability classification compared to DFT as a function of the $E_{hull}$ threshold**. The accuracy is defined as the percentage of the total number of garnets correctly classified as stable/stable, i.e., $E_{hull}$ predicted from the optimized artificial neural network model and DFT are both below/above the threshold. For the mixed garnets, an $E_{hull}$ is calculated for all orderings (20, 7 and 18 orderings per composition for C-, A- and D-mixed, respectively).